\documentclass{tlp}
\pdfoutput=1
\usepackage{stmaryrd}
\usepackage{graphicx}
\usepackage{amssymb}
\usepackage{fancyvrb}
\usepackage{pst-all}
\usepackage{balance}

\newcommand{\field}[1]{\mathbb{#1}}
\newcommand{\N}{\field{N}}
\newcommand{\Z}{\field{Z}}

\newtheorem{definition}{Definition}
\newtheorem{example}{Example}
\newtheorem{theorem}{Theorem}

\newtheorem{proposition}{Proposition}

\begin{document}
\title{Non-termination Analysis of Logic Programs with integer arithmetics}
\author{Dean Voets\thanks{Supported by the Fund for
    Scientific Research - 
	FWO-project G0561-08}
    $~~~~~$ Danny De Schreye \\
Department of Computer Science, K.U.Leuven, Belgium 	\\
 	Celestijnenlaan 200A, 3001 Heverlee			\\
     \{Dean.Voets, Danny.DeSchreye \}@cs.kuleuven.be}

\maketitle

\begin{abstract}
In the past years, analyzers have been introduced to detect classes of non-terminating
queries for definite logic programs. Although these non-termination analyzers have 
shown to be rather precise, their applicability on real-life Prolog programs is limited 
because most Prolog programs use non-logical features. As a first step towards the
analysis of Prolog programs, this paper presents a non-termination condition for 
Logic Programs containing integer arithmetics.
The analyzer is based on our non-termination analyzer presented at ICLP 2009. 
The analysis starts from a class of queries and infers a subclass of non-terminating
ones. In a first phase, we ignore the outcome (success or failure) of the arithmetic
operations, assuming success of all arithmetic calls. In a second phase, we 
characterize successful arithmetic calls as a constraint problem, the solution of 
which determines the non-terminating queries.

\end{abstract}

\textbf{Note:} This article has been published in 
\textit{Theory and Practice of Logic Programming, volume 11, issue 4-5, pages 521-536, 2011}.
 
\section{Introduction}

The problem of proving termination has been studied extensively in Logic Programming. 
Since the early works on termination analysis in Logic Programming, see e.g. \cite{DBLP:journals/jlp/SchreyeD94}, there has been 
a continued interest from the community for the topic. 
Lots of in-language and transformational tools have been developed,
e.g. \cite{Giesl06aprove1.2} and \cite{DBLP:journals/corr/abs-0912-4360}, 
and since 2004, there is an annual Termination 
Competition\footnote{Results are available at http://termcomp.uibk.ac.at/} to compare the current 
analyzers on the basis of an extensive database of logic programs.

In contrast with termination analysis, the dual problem, to detect non-terminating classes
of queries, is a fairly new topic. The development of the first 
and most well-known non-termination analyzer, $NTI$ \cite{nti_06}, was motivated by difficulties in
obtaining precision results for termination analyzers. 
Since the halting problem is undecidable, one way of demonstrating the precision of a termination analyzer 
is with a non-termination analyzer. For $NTI$ it was already shown that for many examples one can partition queries in terminating and non-terminating.
$NTI$ compares the consecutive 
calls in the program using binary unfoldings and proves non-termination by comparing the 
head and body of these binary clauses with a special more general relation.

Recently, in joined work with Yi-Dong Shen, we integrated loop checking into termination
analysis, yielding a very accurate technique to predict the 
termination behavior for classes of queries described using modes \cite{term_prediction}.
Classes of queries are represented as \textit{moded queries}. A moded query consists of
a query and a label, input or output, for each variable in the query. These moded queries
are then evaluated with a \textit{moded SLD-tree} obtained by applying clauses to the 
partially instantiated query and propagating the labels. To guarantee a finite analysis,
this moded SLD-tree is constructed using a complete loop check. After evaluating the moded
query, the analysis predicts the termination behavior of the program for the considered queries
based on the labels and substitutions in the moded SLD-tree.

Motivated by the elegance of this approach and the accuracy of the predictions,
our research focused on defining a non-termination condition based on these moded queries.
In \cite{DBLP:conf/iclp/VoetsS09}, we introduced a non-termination condition identifying paths in a moded SLD-tree that 
can be repeated infinitely often. This approach was implemented in a system called $P2P$, which
proved more accurate than $NTI$ on the benchmark of the termination competition.
An evaluation of the classes of queries not handled
by current approaches lead to considerable improvements in our non-termination analysis. 
These improvements were presented in \cite{VDS10} and implemented in the analyzer $pTNT$.

Both termination and non-termination analyzers have been rather successful in analyzing 
the termination behavior of definite logic programs, but only a few termination analyzers, e.g. \cite{DBLP:conf/lpar/SerebrenikS01a},
and none of the non-termination analyzers handle non-logical features such as arithmetics or cuts,
typically used in practical Prolog programs. 
In this paper, we introduce a technique for proving non-termination of logic programs containing
a subset of the built-in predicates for integer arithmetic, commonly found in Prolog
implementations. 

Given a program, containing integer arithmetics, and a class of queries, described
using modes, we infer a subset of these queries for which we prove existential 
non-termination (i.e. the derivation tree for these queries contains an infinite 
path). The inference and proof are done in two phases. 
In the first phase, non-termination of the logic part of the program is proven by assuming 
that all comparisons between integer expressions succeed. We will show that only a minor adaption
of our technique presented in \cite{DBLP:conf/iclp/VoetsS09} is needed to achieve this. 
In the second phase, given the moded query, integer arguments are identified and constraints
over these arguments are formulated, such that solutions for these constraints correspond
to non-terminating queries.

The paper is structured as follows. In the next section, we introduce some preliminaries 
concerning logic programs, integer arithmetics and we present the symbolic derivation 
trees used to abstract the computation. In Section 3, we introduce our non-termination 
condition for programs containing integer arithmetics. In Section 4, we describe our 
prototype analyzer and some results. Finally, we conclude in Section 5.

\section{Preliminaries}

\subsection{Logic Programming}

We assume the reader is familiar with standard terminology of logic programs, 
in particular with SLD-resolution as described in \cite{Lloyd_foundations}.
Variables are denoted by strings beginning with a capital letter. Predicates, functions 
and constant symbols are denoted by strings beginning with a lower case letter.
We denote the set of terms constructible from a program $P$ by $Term_P$.
Two atoms are called \textit{variants} if they are equal
up to variable renaming. An atom $A$ is \textit{more general} than an atom $B$ and
$B$ is an \textit{instance} of $A$ if there exists a substitution $\theta$ such that $A\theta = B$.

We restrict our attention to definite logic programs.
A logic program $P$ is a finite set
of clauses of the form $H\leftarrow A_1,..., A_n$,
where $H$ and each $A_i$ are atoms.
A goal $G_i$ is a headless clause $\leftarrow A_1,..., A_n$.
A top goal is also called the query. Without loss of generality, we assume
that a query contains only one atom. 

Let $P$ be a logic program and $G_0$ a goal. 
$G_0$ is evaluated by building a \textit{generalized SLD-tree} 
as defined in \cite{term_prediction}, 					
in which each node is represented by $N_i:G_i$ where $N_i$ is the name 
of the node and $G_i$ is a goal attached to the node.
Throughout the paper, we choose to use the best-known \textit{depth-first, left-most} 
control strategy, as is used in Prolog, to select goals and atoms.
So by the \textit{selected atom} in each node $N_i:\leftarrow A_1,..., A_n$,
we refer to the left-most atom $A_1$. For any node $N_i:G_i$, we use 
$A_i^1$ to refer to the selected atom in $G_i$.
Let $A_i^1$ and $A_j^1$ be the selected atoms 
at two nodes $N_i$ and $N_j$, respectively.
$A_i^1$ is an \textit{ancestor} of $A_j^1$
if the proof of $A_i^1$ goes through the proof of $A_j^1$.  

A derivation step is denoted by $N_i:G_i\Longrightarrow_{C} N_{i+1}:G_{i+1}$,
meaning that applying a clause $C$ to $G_i$ produces $N_{i+1}:G_{i+1}$.
Any path of such derivation steps starting at the root node $N_0:G_0$ 
is called a \textit{generalized SLD-derivation}. 

\subsection{Integer arithmetics}

Prolog implementations contain special purpose predicates for handling integer
arithmetics. Examples are $is/2, \geq/2, =:=/2,\ldots$ 

\begin{definition}\label{integer_expressions}
An expression $Expr$ is an \textit{integer expression} if it can be constructed by the following 
recursive definition.
\begin{itemize}
 \item [] $Expr = z \in \Z \mid -Expr \mid Expr+Expr \mid Expr-Expr \mid Expr*Expr$ $\hfill \square$
\end{itemize}
\end{definition}

An atom \verb+"V is Expr"+, with $V$ a free variable and $Expr$ an integer expression, is called 
an \textit{integer constructor}. An atom $Expr1 \circ  Expr2$ is called an 
\textit{integer condition} if $Expr1$ and $Expr2$ are integer expressions and 
$\circ \in \lbrace$\verb+>,>=,=<,<,=:=,=/=+$\rbrace$. 

\subsection{Moded SLD-trees and loop checking}

In \cite{DBLP:conf/iclp/VoetsS09}, classes of queries are represented as \textit{moded queries}. 
Moded queries are partially instantiated queries, in which variables can be labeled as \textit{input}. 
Variables labeled input are called \textit{input variables} and represent arbitrary ground terms.
To indicate that a variable is labeled as input, the name of the variable is underlined.
A query in which no variable is labeled as input is called a \textit{concrete query}.
The set of concrete queries represented by a moded query $Q$ is called the \textit{denotation} of $Q$.

\begin{definition}\label{def:denotation}
Let $Q$ be a query and $\lbrace \underline{I_1},\ldots,\underline{I_n} \rbrace$ its set of
input variables. The \textit{denotation} of $Q$, $Den(Q)$, is defined as:
\begin{itemize}
 \item [] $Den(Q) = \left\lbrace Q\lbrace \underline{I_1} \setminus t_1,\ldots,\underline{I_n} \setminus t_n \rbrace \mid 
	t_i \in Term_P, t_i~is~ground, 1\leq i \leq n \right\rbrace $. $\hfill \square$
\end{itemize}
\end{definition}
Note that the denotation of a concrete query is a singleton containing the query itself. 
Denotations of moded goals and atoms are defined similarly.

A moded query $\leftarrow Q$ is evaluated by constructing a \textit{moded SLD-tree}, representing 
the derivations of the queries in $Den(\leftarrow Q)$. This moded SLD-tree is constructed by applying
SLD-resolution to the query and propagating the labels. An input variable $\underline{I}$ can be 
unified with any term $t \in Term_P$. After unifying $\underline{I}$ and $t$, all variables 
of $t$ will be considered input as well.

\begin{example}\label{example:moded_sld}
Figure \ref{fig:eq_plus_symbolic} shows the moded SLD-tree of the program $eq\_plus$ for the moded query
$\leftarrow eq\_plus(\underline{I},\underline{J},\underline{P})$. This program
is non-terminating for any query in $Den(\leftarrow eq\_plus(\underline{I},\underline{I},0))$
and fails for all other queries in $Den(\leftarrow eq\_plus(\underline{I},\underline{J},\underline{P}))$.
A query fails if its derivation tree is finite, with no path ending with the empty goal.
\begin{verbatim}
eq_plus(I,J,P):- eq(I,J), plus(P,I,In), eq_plus(In,J,P).
eq(A,A).     plus(0,B,B).     plus(s(A),B,s(C)):- plus(A,B,C).
\end{verbatim}

\begin{figure}[htp]
\centering
\includegraphics[width=70ex]{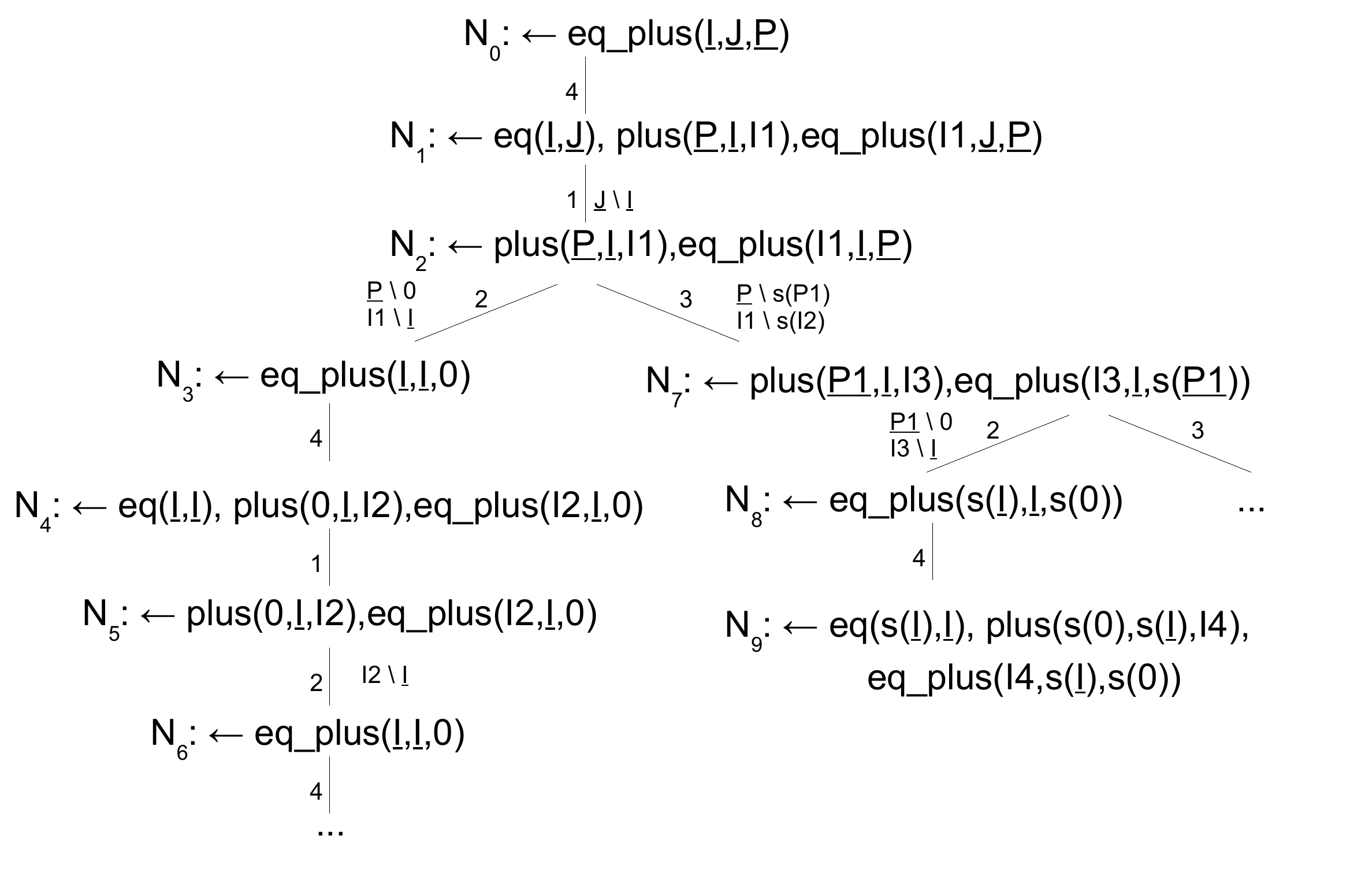}
\vspace*{-15pt}
\caption{Moded SLD-tree $eq\_plus$}\label{fig:eq_plus_symbolic}
\vspace*{-15pt}
\end{figure}

Substitutions on input variables express conditions for the clause to be applicable. 
The edge from node $N_2$ to $N_3$ shows that clause two is applicable if the concrete 
term denoted by $\underline{P}$ can be unified with $0$. The substitution, 
$I1 \setminus \underline{I}$, shows that applying this clause
unifies $I1$ with the term corresponding to $\underline{I}$.

Every derivation in a moded SLD-tree for a query $\leftarrow Q$ corresponds to a
concrete derivation for a subclass of $Den(\leftarrow Q)$. The subclass of queries for
which a derivation to node $N_i$ is applicable is obtained by applying all substitutions
on input variables from $N_0$ to $N_i$.
Our condition of \cite{DBLP:conf/iclp/VoetsS09} proves non-termination for every query for which the derivation to
$N_3$ is applicable. The substitutions on input variables in the derivation to $N_3$ are
$\underline{J}\setminus \underline{I}$ and $\underline{P} \setminus 0$. Applying these
to the query proves non-termination for the queries in $Den(\leftarrow eq\_plus(\underline{I},\underline{I},0))$.
$\hfill \square$
\end{example}

As in the example, moded SLD-trees are usually infinite. To obtain a finite analysis, 
a complete loop check is applied during the construction of the tree. As in our
previous works, \cite{DBLP:conf/iclp/VoetsS09} \cite{VDS10}, we use the complete loop check 
\textit{LP-check}, \cite{shen_dynamic_approach}.
Without proof, we state that this loop check can also be used for moded SLD-trees and refer
to \cite{shen_dynamic_approach} for more information.


\begin{example}
In Figure \ref{fig:eq_plus_symbolic}, LP-check cuts clause 4 at node $N_6$ and clause 3 at node $N_7$.
$\hfill \square$
\end{example}

Combined with the loop check, a moded SLD-tree can be considered a light-weight alternative
to an abstract interpretation for mode analysis.

\section{Non-termination analysis for programs with integer arithmetics}

In this section, we introduce a non-termination condition for programs containing 
integer arithmetics. To abstract the computations for the considered queries, the
moded SLD-tree of \cite{term_prediction} is used, with some modifications to handle integer constructors
and integer conditions. LP-check ensures finiteness of the tree and detects paths that
may correspond to infinite loops. For every such path, two analyses are combined
to identify classes of non-terminating queries.

In the first phase, an adaption of the non-termination condition of 
\cite{DBLP:conf/iclp/VoetsS09} detects a class of queries such that each query is
non-terminating or fails due to the evaluation of an integer condition such as $>/2$.
This class of queries is a moded query with an additional integer label for variables
representing unknown integers. In the second phase, the class of queries is restricted
to a class of non-terminating queries by formulating additional constraints on the integer 
variables of the moded query. 
To prove that this class of non-terminating queries is not 
empty, these constraints over unknown integers are transformed to constraints over the 
natural numbers and solved by applying well-known techniques from termination analysis.
Then we try to solve these constraints by transforming them to constraints over the 
natural numbers and applying well-known techniques on them.

\subsection{Moded SLD-tree for programs with integer arithmetics}

The first step of the extension is rather straightforward.
The extensions to the moded SLD-tree of \cite{term_prediction} are limited to the
introduction of the label \textit{integer variable} and additional transitions to handle 
integer constructors and integer conditions. Integer variables are also input variables 
and will also be represented by underlining the name of the variable. An integer constructor, i.e. $is/2$,
is applicable if the first argument is a free variable and the second argument is an integer 
expression. The application of an integer constructor labels the free variable as an
integer variable. An integer condition, e.g. $\geq/2$, is applicable if 
both arguments are integer expressions. Since integer variables denote unknown integers, 
integer expressions are allowed to contain integer variables. Applications of integer
constructors and integer conditions in the moded SLD-tree are denoted by derivation
steps $N_i:G_i\Longrightarrow_{cons} N_{i+1}:G_{i+1}$ and 
$N_i:G_i\Longrightarrow_{cond} N_{i+1}:G_{i+1}$, respectively.

\begin{example}\label{example:count_to}
The following program, $count\_to$, is a faulty implementation of a predicate generating
the list starting from 0 up to a given number. The considered class of queries is 
represented by the moded query $\leftarrow count\_to(\underline{N},L)$ with $\underline{N}$ 
an integer variable.

\begin{verbatim}
count_to(N,L):- count(0,N,L).       count(N,N,[N]).
count(M,N,[M|L]):- M > N, M1 is M+1, count(M1,N,L).
\end{verbatim}

In the last clause, the integer condition should be \verb+M < N+ instead of \verb+M > N+.
Due to this error, the program:
\begin{itemize}
\item fails for the queries for which $\underline{N}>0$ holds,
\item succeeds for $\leftarrow count\_to(0,L)$,
\item loops for the queries for which $\underline{N} < 0$ holds.
\end{itemize}

\begin{figure}[htp]
\centering
\includegraphics[width=60ex]{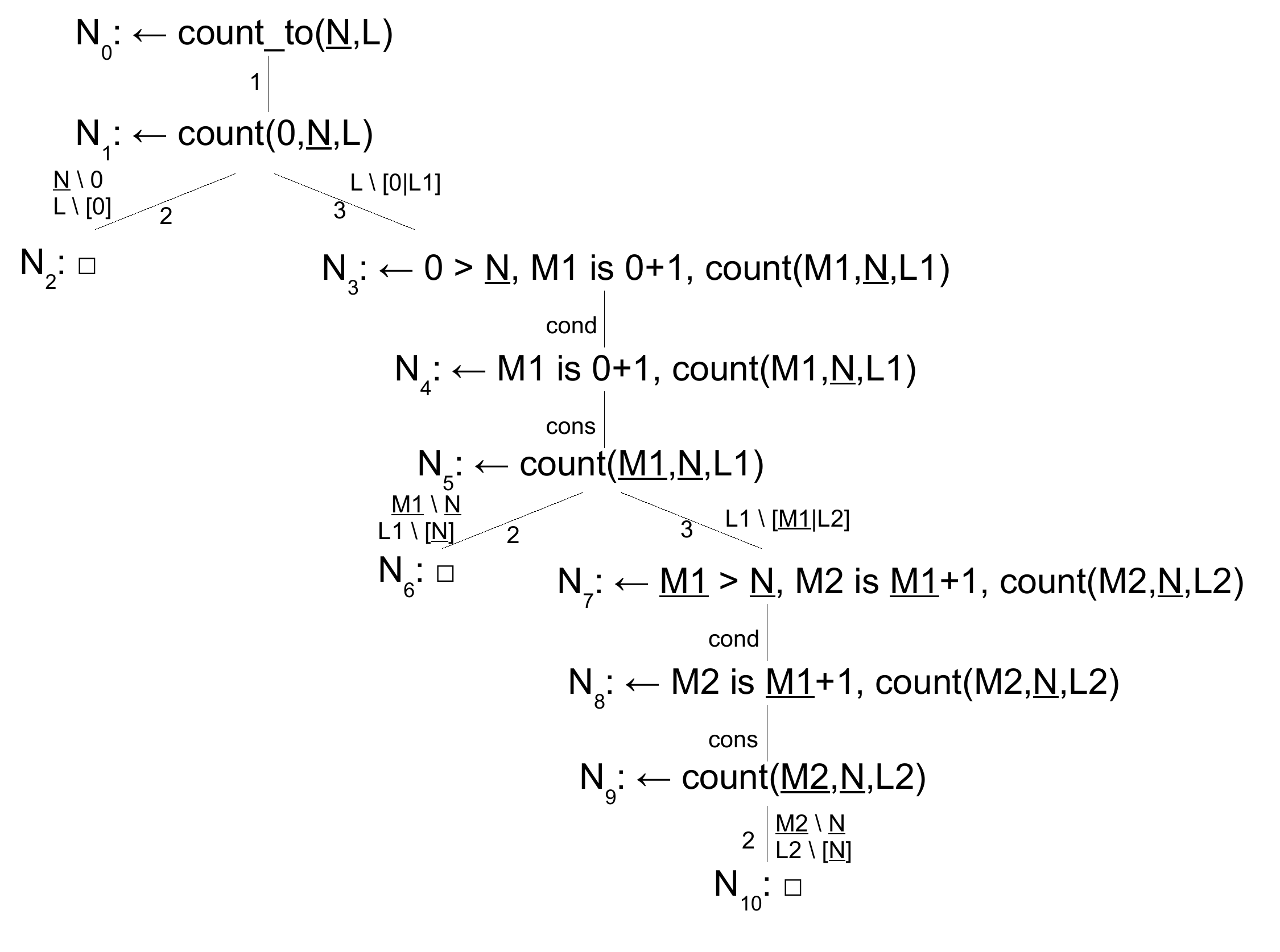}
\vspace*{-15pt}
\caption{Moded SLD-tree $count\_to$}\label{fig:count_to}
\vspace*{-10pt}
\end{figure}

Figure \ref{fig:count_to} shows the moded SLD-tree for the considered query, constructed
using LP-check. LP-check cuts clause 3 at node $N_9$.
$\hfill \square$
\end{example}

Note that by ignoring the possible values for the integer variables when constructing the tree,
some derivations in it may not be applicable to any considered query. For example the refutations 
at nodes $N_6$ and $N_{10}$ in the previous example cannot be reached by the considered queries.

\subsection{Adapting the non-termination condition}

In \cite{DBLP:conf/iclp/VoetsS09}, programs are shown to be non-terminating for a moded query, by proving that
a path in the moded SLD-tree can be repeated infinitely often. Such a path, from a node 
$N_b$ to a node $N_e$, is identified based on three properties. The path should be applicable, 
independent from the concrete terms represented by the input variables. Therefore, the
first property states that no substitutions on input variables may occur between $N_b$ 
and $N_e$. The second property states that the selected atom of $N_b$ -- i.e. $A_b^1$ -- has to
be an ancestor of $A_e^1$. These two properties prove that the sequence of clauses in 
the path from $N_b$ to $N_e$ is applicable to any goal with a selected atom from $Den(A_b^1)$. 
Therefore, non-termination is proven by requiring that $Den(A_e^1)$ is a subset of $Den(A_b^1)$. 
This property can be relaxed by requiring that each atom in $Den(A_e^1)$ is more general
than some atom in $Den(A_b^1)$. If this is the case, $A_e^1$ is called 
\textit{moded more general} than $A_b^1$. For definite logic programs, these three 
properties imply non-termination.

\begin{definition}
Let $A$ and $B$ be moded atoms. $A$ is \textit{moded more general} than $B$ if
\begin{itemize}
 \item [] $ \forall I \in Den(A),~ \exists J \in Den(B): I \textit{ is more general than } J$.$\hfill \square$
\end{itemize}
\end{definition}

\begin{example}
In Figure \ref{fig:eq_plus_symbolic}, the path from $N_3$ to $N_6$ 
satisfies these properties. The ancestor relation holds. There are no substitutions on
input variables in the path. Finally, the selected atoms are identical and therefore denote
the same concrete atoms.
$\hfill \square$
\end{example}

The following proposition provides a practical sufficient condition to verify
whether the moded more general relation holds.
\begin{proposition}[Proposition 1 of \cite{DBLP:conf/iclp/VoetsS09}
]\label{prop:mmg}
Let $A$ and $B$ be moded atoms. Let $A_1$ and $B_1$ be renamings of these atoms
such that they do not share variables. $A$ is moded more general than $B$ if
$A_1$ and $B_1$ are unifiable with most general unifier $\lbrace V_1\setminus t_1,\ldots,
V_n \setminus t_n \rbrace$, $t_i \in Term_P$, $1 \leq i, \leq n$, such that for each binding $V_i \setminus t_i$, either:
\begin{itemize}
\item $V_i \in Var(B_1)$ and $V_i$ is labeled as input, or
\item $V_i \in Var(A_1)$, $V_i$ is not labeled as input and no variable of $Var(t_i)$ 
is labeled as input. $\hfill \square$
\end{itemize}
\end{proposition}

As stated, we want to prove that every query in the denotation of the considered 
moded query is either non-terminating or terminates due to the evaluation of an
integer condition. To achieve this, we need to guarantee that integer constructors
are repeatedly evaluated with a free variable and an integer expression as arguments 
and that integer conditions are repeatedly evaluated with integer expressions as
arguments. Proposition \ref{prop:mmg} already implies that the first argument of all
integer constructors are free variables in the subsequent iterations of the loop.

To prove the repeated behavior on integer constructors and integer expressions stated
above, the \textit{integer-similar to} relation is defined. Intuitively, given some 
loop  in the computation, if an atom at the end of the loop is integer-similar to 
an atom at the start of the loop, then it will provide the required integer expressions
to the first atom. First, we introduce positions to identify subterms and a 
function to obtain a subterm from a given position.

\begin{definition}\label{def:func_subterm}
Let $L$ be a list of natural numbers, called a \emph{position}, and $A$ a moded atom or term. 
The function \emph{subterm(L,A)} returns the
subterm obtained by:
\begin{itemize}
\item if $L = [I]$ and $A=f(A_1,\ldots,A_I,A_{I+1},\ldots,A_n)$ then $subterm(L,A) = A_I$
\item else if $L=[I|T]$ and $A=f(A_1,\ldots,A_I,A_{I+1},\ldots,A_n)$ then $subterm(L,A) = subterm(T,A_I)$ $\hfill \square$
\end{itemize}
\end{definition}

An atom $A$ is integer-similar to an atom $B$ if it has integer expressions on all 
positions corresponding to integer expressions in $B$.
\begin{definition}
Let $A$ and $B$ be moded atoms. $A$ is \textit{integer-similar to} $B$ if for every 
integer expression $t_B$ of $B$, with $subterm(L,B) = t_B$, there exists an 
integer expression $t_A$ of $A$, with $subterm(L,A) = t_A$. $\hfill \square$
\end{definition}

\begin{example}
\begin{itemize}
 \item $count(0,\underline{N},L)$ is integer-similar to $count(\underline{M},\underline{N},L)$
 \item $count(\underline{M},\underline{N},L)$ is integer-similar to $count(0,\underline{N},L)$
 \item $count(\underline{M} + 1,\underline{N},L)$ is integer-similar to $count(\underline{M},\underline{N},L)$
 \item $count(\underline{M},\underline{N},L)$ is not integer-similar to $count(\underline{M}+1,\underline{N},L)$
\end{itemize}
Note that the last one is a counterexample because $count(\underline{M}+1,\underline{N},L)$ has
integer expressions on $[1,1]$ and $[1,2]$, while $count(\underline{M},\underline{N},L)$ does not
have any subterms on these positions. 
$\hfill \square$
\end{example}

\begin{theorem}\label{th:analysis1}
Let $N_b$ and $N_e$ be nodes in a moded SLD-tree for a moded query $Q$. Let
$Q'$ be the moded atom obtained by applying to $Q$ all substitutions on input variables 
from $N_0$ to $N_b$. Every query in $Den(Q')$ is either non-terminating or terminates
due to the evaluation of an integer condition if the following properties hold:
\begin{itemize}
\item $A_b^1$ is an ancestor of $A_e^1$
\item no substitutions on input variables occur from $N_b$ to $N_e$
\item $A_e^1$ is moded more general than $A_b^1$
\item $A_e^1$ is integer-similar to $A_b^1$ $\hfill \square$
\end{itemize}
\end{theorem}

\begin{example}\label{example:mmg_adaption}
The path between nodes $N_5$ and $N_9$ in Figure \ref{fig:count_to} satisfies the conditions
of Theorem \ref{th:analysis1}. There are no substitutions on input variables from $N_0$ to 
$N_5$ and thus, every query in $Den(\leftarrow count\_to(\underline{N},L))$ is either non-terminating
or fails due to the evaluation of an integer condition. Note that although $\leftarrow count\_to(0,L)$
has a succeeding derivation to $N_2$, its derivation to $N_9$ fails due to the 
integer condition $0 > \underline{N}$.
$\hfill \square$
\end{example}

To verify the last property automatically, we strengthen Proposition \ref{prop:mmg} to 
imply both the moded more general relation and the integer-similar to relation.

\begin{proposition}\label{prop:mmg_int_ins}
Let $A$ and $B$ be moded atoms. Let $A_1$ and $B_1$ be renamings of these atoms
such that they do not share variables. $A$ is moded more general than $B$ and $A$ is integer-similar to $B$, if
$A_1$ and $B_1$ are unifiable with most general unifier $\lbrace V_1\setminus t_1,\ldots,
V_n \setminus t_n \rbrace$, such that for each binding $V_i \setminus t_i$, $1\leq i \leq n$, either:
\begin{itemize}
\item $V_i \in Var(B_1)$ and $V_i$ is labeled as integer and $t_i$ is an integer expression, or
\item $V_i \in Var(B_1)$ and $V_i$ is labeled as input but not as integer variable, or
\item $V_i \in Var(A_1)$, $V_i$ is not labeled as input, no variable of $Var(t_i)$ 
is labeled as input and $t_i$ does not contain integers. $\hfill \square$
\end{itemize}
\end{proposition}

\begin{example}
Since the selected atoms of nodes $N_5$ and $N_9$ in Figure \ref{fig:count_to} are
variants, Proposition \ref{prop:mmg_int_ins} holds. $\hfill \square$
\end{example}

\subsection{Generating the constraints on the integers of the query}

In this subsection, we
introduce the constraints on the integer variables of the moded query, identifying values
for which all integer conditions in the considered derivations succeed. These constraints
consist of reachability constraints, identifying queries for which the derivation up till the last node 
is applicable, and an implication proving that the integer conditions will also succeed in 
the following iterations.

\begin{example}\label{example:count_to_int_cons}
As a first example, we introduce the constraints for the path between $N_5$ and $N_9$ in 
the moded SLD-tree of $count\_to$ in Figure \ref{fig:count_to}. For this path, 
Theorem \ref{th:analysis1} holds and thus every query denoted by 
$\leftarrow count\_to(\underline{N},L)$ is either non-terminating or terminates due to 
an integer condition.

To restrict the class of considered queries to those for which the derivation to
$N_9$ is applicable, all integer conditions in the derivation are expressed in terms
of the integers of the query, yielding $0 > \underline{N}$ and $0 + 1 > \underline{N}$.

For this program and considered class of queries, the condition $0 > \underline{N}$ implies that 
the derivation is applicable until node $N_9$. The following implication states 
that if the condition of node $N_7$ holds for any two values $M$ and $N$, 
then it also holds for the values of the next iteration. 
$$\forall M,N \in \Z: M > N \Longrightarrow M+1>N$$
This implication is correct and thus proves non-termination for the considered queries
if the precondition holds in the first iteration. This is the case for all queries
in $Den(\leftarrow count\_to(\underline{N},L))$ with $0 > \underline{N}$ since the value
corresponding to $M$ in the first iteration is $0$ and the value corresponding to $N$ is $\underline{N}$.
This proves non-termination of all considered queries for which $0 > \underline{N}$.
$\hfill \square$
\end{example}

In the following example, applicability of the derivation does not imply non-termination.
To detect a class of non-terminating queries, a domain constraint is added to the pre- and 
postcondition of the implication.
\begin{example}\label{example:constants_nt_cond}
\begin{verbatim}
constants(I,J):- I =:= 2, In is J*2, Jn is I-J, constants(In,Jn).
\end{verbatim}
The clause in \textit{constants} is applicable to any goal with
$constants(2,\underline{J})$ as selected atom, with $\underline{J}$ an integer variable. 
Since the first argument in the next iteration is the value corresponding to $\underline{J}*2$, 
only goals with the selected atom $constants(2,1)$ are non-terminating for this program.

Since applicability of the derivation does
not imply non-termination, a similar implication as in the previous example is false, 
$\forall I,J \in \Z: I=2 \Longrightarrow J*2 = 2$. To overcome this, a constraint is
added to the pre- and post-condition of this implication, restricting the considered values of 
$\underline{J}$ to an unknown set of integers, called its \textit{domain}.
$$\exists Dom_j \subset \Z, \forall I,J \in \Z: I=2, J \in Dom_j \Longrightarrow J*2 = 2, I-J \in Dom_j$$
The resulting implication is true for $Dom_j = \lbrace 1 \rbrace$. By requiring that
the considered moded query satisfies both the reachability constraint and the additional
constraint in the pre-condition, the non-terminating query $\leftarrow constants(2,1)$ is
obtained.
$\hfill \square$
\end{example}

All information needed to construct these constraints can be obtained from the
moded SLD-tree.

\begin{definition}
Let $C$ be an integer condition or expression and $N_i$ and $N_j$ two nodes in a 
moded SLD-tree $D$. Let $Cons$ be the set of all integer constructors occurring as
selected atom in a node $N_p~(i \leq p \leq j)$ in $D$.

The function \emph{$apply\_cons(C,N_i,N_j)$} returns the integer 
condition or expression obtained by exhaustively applying $\underline{I}\setminus Expr$ to $C$, 
for any $\underline{I} ~is~ Expr \in Cons$. 
$\hfill \square$
\end{definition}

The constraints guaranteeing a derivation to $N_j$ to be applicable, can be obtained 
using $apply\_cons(Cond,N_0,N_i)$ for any integer condition $Cond$ in a node $N_i$ in the considered derivation. 
For a path from $N_b$ to $N_e$, the precondition of the implication is obtained using $apply\_cons(Cond,N_b,N_i)$, for each
condition $Cond$ in a node $N_i$ between nodes $N_b$ to $N_e$ and universally quantifying the integer variables of $N_b$.
\begin{example}\label{example:apply_cons}
The derivation to $N_9$ in Figure \ref{fig:count_to}, contains integer conditions
in nodes $N_3$ and $N_7$. These are expressed on the integer variable of the query, $\underline{N}$,
using $apply\_cons$.
\begin{itemize}
 \item $apply\_cons(0>\underline{N},N_0,N_3) = 0 > \underline{N}$
 \item $apply\_cons(\underline{M1}>\underline{N},N_0,N_7) = 0 + 1 > \underline{N}$
\end{itemize}
To obtain the precondition of the implication, the integer condition in $N_7$ is expressed 
in terms of the integer variables of $N_5$. 
\begin{itemize}
 \item $apply\_cons(\underline{M1}>\underline{N},N_5,N_7) = \underline{M1} > \underline{N}$
\end{itemize}
Universally quantifying these variables yields the precondition. $\hfill \square$
\end{example}
To obtain the consequence of the implication for a path from $N_b$ to $N_e$, 
one first replaces the integer variables of $N_b$ in the precondition by 
the corresponding integer variables of $N_e$. Then, $apply\_cons$ is used 
to express the consequence in terms of the values in the previous iteration.

\begin{definition}
Let $LHS$ be the precondition of an implication, consisting of integer conditions and
constraints of the form $I \in Dom_I$. Let $N_i$ and $N_j$ be two nodes in a moded 
SLD-derivation such that all integer variables in $LHS$ are in $A_i^1$ and let 
$\underline{I_1},\ldots,\underline{I_n}$ be all integer variables of $A_i^1$.

If there exist subterms of $A_j^1$, $t_1,\ldots,t_n$, such that $\forall L: subterm(L,A_i^1)=\underline{I_p} \Longrightarrow 
subterm(L,A_j^1)=t_p, 1 \leq p \leq n$, then \emph{$replace(LHS,N_i,N_j)$} is
obtained by applying $\lbrace \underline{I_1} \setminus t_1, \ldots, \underline{I_n} \setminus t_n\rbrace$
to all constraints in $LHS$.
$\hfill \square$
\end{definition}
\begin{example}
In Example \ref{example:apply_cons}, we generated the precondition of the implication, $\underline{M1} > \underline{N}$.
To obtain the consequence, $replace(\underline{M1} > \underline{N},N_5,N_9)$ is applied,
yielding $\underline{M2} > \underline{N}$. Then, the integer variable of $N_9$, $\underline{M_2}$,
is expressed in terms of the integer variables of $N_5$ using 
$apply\_cons(\underline{M2} > \underline{N},N_5,N_9)=\underline{M_1}+1 > \underline{N}$.

Adding the domains to the pre- and postcondition yields the desired implication: 
$\exists Dom_N, Dom_{M1} \subset \Z, \forall N,M1 \in \Z: M1 > N,~N \in Dom_N,~M1 \in Dom_{M1} \Longrightarrow$ 
\\$~~~~~~~~~M1+1 > N,~N \in Dom_N,~M1+1 \in Dom_M$ $\hfill \square$
\end{example}
Adding these constraints to the class of queries detected by Theorem \ref{th:analysis1},
yields a class of non-terminating queries.

\subsection{Proving that the constraints on integers are solvable}
The previous subsection introduced constraints, implying that all integer conditions
in a considered derivation succeed. In this subsection, we introduce a technique to 
check if these constraints have solutions, using a constraint-based approach. Symbolic 
coefficients represent values for the integers in the query and domains in the implication, 
for which the considered path is a loop. After these coefficients are introduced, the
implication is transformed into a set of equivalent implications over natural numbers.
These implications can then be solved automatically in the constraint-based approach,
based on Proposition 3 of \cite{DBLP:journals/corr/abs-0912-4360}.	

\begin{proposition}[Proposition 3 of \cite{DBLP:journals/corr/abs-0912-4360}]\label{prop:rem_imp}
Let $prem$ be a polynomial over $n$ variables and $conc$ a polynomial over 1 variable, both with 
natural coefficients, where $conc$ is not a constant. 
Moreover, let $p_1,\ldots,p_{n+1},q_1,\ldots,q_{n+1}$ be arbitrary polynomials 
with integer coefficients\footnote{Proposition 3 in \cite{DBLP:journals/corr/abs-0912-4360} 
uses natural coefficients, but the proposition also holds for polynomials with integer 
coefficients.} over the variables $\overline{X}$. If
$$\forall \overline{X} \in \N: conc(p_{n+1})-conc(q_{n+1})-prem(p_1,\ldots,p_n)+
prem(q_1,\ldots,q_n) \geq 0$$
is valid, then
$\forall \overline{X} \in \N: p_1 \geq q_1, \ldots,p_n\geq q_n \Longrightarrow p_{n+1}\geq q_{n+1}$
is also valid. $\hfill \square$
\end{proposition}

\subsubsection{Introducing the symbolic coefficients.}

To represent half-open domains in the implication by symbolic coefficients, the domains are
described by two symbolic coefficients, one upper or lower limit and one for the direction.
Constraints of the form $Exp \in Dom_I$ in the implication, are replaced by constraints 
of the form $d_I * Exp \geq d_I* c_I$ with $d_I$ either $1$ or $-1$, describing 
the domain $\lbrace c_I, c_I-1, \ldots\rbrace$ for $d_I=-1$ and $\lbrace c_I, c_I+1, \ldots\rbrace$ for $d=1$. 
The values to be inferred for the integers of the query should satisfy the precondition 
of the implication. Off course, the symbolic coefficients $c_I$ should also be consistent
with the values of the integers in the query.

\begin{example}
In Example \ref{example:count_to_int_cons}, we introduced constraints on the integer
variable $\underline{N}$, $0 > \underline{N}$ and $0 + 1 > \underline{N}$, 
proving non-termination for queries in $Den(\leftarrow count\_to(\underline{N},L))$.
By convention, we denote the symbolic coefficients as constants. For the integer variable 
$\underline{N}$, we introduce the symbolic coefficient $n$.

The implication introduced in Example \ref{example:count_to_int_cons}, for the path from $N_5$ to $N_9$ in 
Figure \ref{fig:count_to}, does not contain constraints on the domains. When adding these 
constraints to the pre- and postcondition, we obtain the following implication.
\begin{itemize}
 \item [] $\forall M,N \in \Z: ~M > N, ~N \in Dom_N, ~M \in Dom_M \Longrightarrow $ 
\\$~~~~~~~~~~~M+1>N, ~N \in Dom_N, ~M+1 \in Dom_M$
\end{itemize}

Representing these domains by symbolic coefficients yields the following implication.
\begin{itemize}
 \item [] $\forall M,N \in \Z: ~M > N, ~d_N * N \geq d_N * c_N, ~d_M * M \geq d_M * c_M \Longrightarrow $ 
\\$~~~~~~~~~~~M+1>N, ~d_N * N \geq d_N * c_N, ~d_M * (M+1) \geq d_M * c_M$
\end{itemize}

To guarantee that the precondition succeeds for the considered derivation, $c_M$ and $c_N$ 
are required to be the values for $\underline{M}$ and $\underline{N}$ in node $N_5$. Combining
these constraints implies non-termination for the query $\leftarrow count\_to(n,L)$, for 
which the following constraints are satisfied with some unknown integers $c_N,c_M,d_N$ and $d_M$.
\begin{itemize}
\item [(1)] $0>n,~0+1>n$ to guarantee applicability of the derivation
\item [(2)] $c_N = n, ~c_M = 0+1$ to guarantee that the precondition holds
\item [(3)] $d_N = 1 \lor d_N = -1, ~d_M = 1 \lor d_M = -1$,
\item [(4)] $\forall M,N \in \Z: M > N, d_N * N \geq d_N * c_N, d_M * M \geq d_M * c_M \Longrightarrow $ 
\\$~~~~~~~~~~~M+1>N, d_N * N \geq d_N * c_N, d_M * (M+1) \geq d_M * c_M$ to prove that the condition succeeds infinitely often.
\end{itemize}
Due to the implication, $d_M$ has to be $1$. $d_N$ can be either $1$ or $-1$. $\hfill \square$
\end{example}

To be able to infer singleton domains, we allow the constant describing
the direction of the interval to be $0$. If in such a constant $d_I$ is zero, the constraints
on the domain are satisfied trivially because they simplify to $0 \geq 0$. 
To guarantee that the domain is indeed a singleton when $d_I$ is inferred to be zero,
a constraint of the form $(1-d_I^2)Exp=(1-d_I^2)*c_I$ is added to the postcondition
for every constraint $d_I * I \geq d_I * c_I$.
This constraint is trivially satisfied for half-open domains and proves that $\lbrace c_I \rbrace$
is the domain in the case that $d_I = 0$.

\begin{example}
In Example \ref{example:constants_nt_cond}, we introduced constraints on the integer variables
$\underline{I}$ and $\underline{J}$, proving non-termination for queries in 
$Den(\leftarrow constants(\underline{I},\underline{J}))$. Introducing symbolic coefficient $i$ and 
$j$ for the integers of the query and for the domains of $\underline{I}$ and $\underline{J}$, yields
the following constraints.
\begin{enumerate}
\item [(1)] $i = 2$ to guarantee applicability of the derivation
\item [(2)] $c_I = i, ~c_J = j$ to guarantee that the precondition holds
\item [(3)] $d_I \leq 1, ~d_I \geq  -1, ~d_J \leq 1, ~d_J \geq -1$,
\item [(4)] $\forall I,J \in \Z: I=2, ~d_I * I \geq d_I * c_I, ~d_J * J \geq d_J * c_J \Longrightarrow $ 
\\$~~~~~J*2=2, ~d_I * (J*2) \geq d_I * c_I, (1-d_I^2)*(J*2) = (1-d_I^2)*c_I, $
\\$~~~~~d_J * (I-J) \geq d_J * c_J, (1-d_J^2)*(I-J) = (1-d_J^2)*c_J$
\end{enumerate}
The implication in $(4)$ can only be satisfied with $d_J$ equal to zero. $\hfill \square$
\end{example}

\subsubsection{To implications over the natural numbers}
The symbolic coefficients to be inferred which represent the domains, allow to transform 
the implication over $\Z$ to an equivalent implication over $\N$.
\begin{itemize}
\item  for $d_I = 1$, any integer in $\lbrace c_I,~c_I+1,~\ldots\rbrace$ that satisfies the 
	precondition is in $\lbrace c_I+d_I*N \mid N \in \N \rbrace$
\item  for $d_I = -1$, any integer in $\lbrace c_I,~c_I-1,~\ldots\rbrace$ that satisfies the 
	precondition is in $\lbrace c_I+d_I*N \mid N \in \N \rbrace$
\item  for $d_I = 0$, any integer in $\lbrace c_I \rbrace$ that satisfies the 
	precondition is in $\lbrace c_I+d_I*N \mid N \in \N \rbrace$
\end{itemize}

Therefore, we obtain an equivalent implication over the natural numbers by replacing
each integer $I$ by its corresponding expression $c_I+d_I*N$ and replacing the universal
quantifier over $I$ by a quantifier over $N$.

\subsubsection{Automation by a translation to diophantine constraints}
To solve the resulting constraints, we use the approach of \cite{DBLP:journals/corr/abs-0912-4360}.
Constraints of the form $A =:= B$ in the implication, are replaced by the conjunction $A\geq B,~B\geq A$. 
Constraints of the form $A =/= B$, yield two disjunctive cases. One obtained by replacing the $=/=$ 
in the pre- and postcondition by $>$ and one obtained by replacing it by $<$. 
The other conditions -- i.e. $>,<$ and $\leq$ -- are transformed into $\geq$-constraints in the obvious way.
Implications with only one consequence are obtained by creating one implication for each consequence,
with the pre-condition of the original implication.

The resulting implications allow to apply Proposition \ref{prop:rem_imp}. These 
inequalities of the form, $p\geq0$, are then transformed into a set of \textit{diophantine 
constraints}, i.e. constraints without universally quantified variables, by requiring that all
coefficients of $p$ are non-negative. As proposed in \cite{DBLP:journals/corr/abs-0912-4360},
the resulting diophantine constraints are then transformed into a SAT-problem. The 
constraints are then proven to have solutions by a SAT solver by inferring one possible solution.

\section{Evaluation}

We have implemented our analysis and integrated it within our existing non-termination 
analyzer $pTNT$. The analyzer can be downloaded from 
\\http://www.cs.kuleuven.be/\~{}dean/iclp2011.html.
We tested our analysis on a benchmark of 16 programs similar to those in the paper. 
These programs are also available online. 
To solve the resulting SAT-Problem, MiniSat \cite{ES03} is used.

\begin{table}[htb]
 \begin{center} 
{\scriptsize
\begin{tabular}{lcccc}
$ $		& linear-class, 3 bits 	&	linear-class, 4 bits		& max2-class, 3 bits		& max2-class, 4 bits	\\
\hline
count\_to	&	$+$		&	$+$			&	$+$		&	$+$	\\
constants	&	$+$		&	$+$			&	$+$		&	$OS$	\\
int1		&	$+$		&	$+$			&	$+$		&	$+$	\\
int2		&	$+$		&	$+$			&	$+$		&	$+$	\\
int3		&	$+$		&	$+$			&	$+$		&	$OS$	\\
int4		&	$+$		&	$+$			&	$+$		&	$OS$	\\
int5		&	$+$		&	$+$			&	$+$		&	$OS$	\\
int6		&	$+$		&	$+$			&	$+$		&	$OS$	\\
int7		&	$+$		&	$+$			&	$OS$		&	$OS$	\\
int8		&	$+$		&	$+$			&	$OS$		&	$OS$	\\
int9		&	$-$		&	$+$			&	$OS$		&	$OS$	\\
int10		&	$-$		&	$-$			&	$+$		&	$OS$	\\
int11		&	$-$		&	$+$			&	$-$		&	$OS$	\\
int12		&	$-$		&	$+$			&	$-$		&	$OS$	\\
int13		&	$+$		&	$+$			&	$+$		&	$+$	\\
int14		&	$+$		&	$+$			&	$+$		&	$OS$	\\
\end{tabular}
}
 \end{center}
\caption{An overview of the experiments}\label{Table:evaluation}
\vspace*{-10pt}
\end{table}
\normalsize


We experimented with different bit-sizes in the translation to SAT and different 
classes of functions for the $prem$ functions in Proposition \ref{prop:rem_imp}. 
As $conc$ functions, the identity function was used.
Table \ref{Table:evaluation} shows the results for the considered settings, $+$ 
denotes that non-termination is proven successfully, $-$ denotes that 
non-termination could not be proven and $OS$ denotes that the computation 
went out of stack. The considered settings are 3 and 4 as bit-sizes and $linear$ and 
$max2$ as forms for the symbolic $prem$-functions. The $linear$ class is 
a weighted sum of each argument. The $max2$ class contains a weighted term 
for each multiplication of two arguments. The analysis time is between $1$ and $20$ 
seconds for all programs and settings.

Table \ref{Table:evaluation} shows non-termination can be proven for any program of 
the benchmark when choosing the right combination of parameters, but no setting 
succeeds in proving non-termination for all programs. Programs $int9$ and $int12$ 
require a constant that cannot be represented with bit-size 3. Linear prem-functions 
cannot prove non-termination for $int10$. However, the setting with 4 as a bit-size
and $max2$ as class of $prem$-function usually fails, because these settings cause 
an exponential increase in memory use during the translation to SAT.


\section{Conclusion}

In this paper we introduced a technique to detect classes of non-terminating queries for 
logic programs with integer arithmetic. The analysis starts with a given program and class
queries, specified using modes, and detects subclasses of non-terminating queries.
First, the derivations for the given class of queries are abstracted by building a 
moded SLD-tree \cite{term_prediction} with additional transitions to handle integer
arithmetic. Then, this moded SLD-tree is used to detect subclasses of non-terminating
queries in two phases. In the first phase, we ignore the conditions over integers, 
e.g. $>/2$, and detect paths in the moded SLD-tree that correspond to infinite derivations
if all conditions on integers in those derivations succeed. For every such path, the
corresponding subclass of queries is generated.
In the second phase, the obtained classes of queries are restricted
to classes of non-terminating queries, by formulating constraints implying that all
conditions on integers will succeed. These constraints are then solved by transforming 
them into a SAT problem.

We implemented this approach in our non-termination analyzer $pTNT$ and evaluated it on
small benchmark of non-terminating Prolog programs with integer arithmetic. 
The evaluation shows that the proposed technique is rather powerful, but also that 
the parameters in the transformation to SAT must be chosen carefully to avoid 
excessive memory use. For future work, we plan to improve the efficiency by using SMT solvers.
\vspace*{-10pt}
\paragraph{Acknowledgment} We thank the referees for their useful and constructive comments.
\vspace*{-20pt}
\bibliography{prolog.bib}
\end{document}